**Manuscript title:** Metamaterial radiofrequency lens for magnetic resonance imaging

**Manuscript type:** Original research

**Authors:**

Manuel J. Freire[1], Ph.D., Associate Professor of Physics

Ricardo Marqués[1], Ph.D., Associate Professor of Physics

Lukas Jelinek[1], Ph.D., Postdoctoral Researcher in Physics

[1]Department of Electronics and Electromagnetism, University of Seville

Facultad de Física, Avenida Reina Mercedes s/n. 41012 Sevilla (Spain)

Eduardo Gil[2], MD

Francisco Moya[2], MD

[2]Centro PET Cartuja

c/ Torricelli, 22. Isla de la Cartuja. 41092 Sevilla (Spain)

**Advances in Knowledge:** This manuscript shows a new electromagnetic device that behaves like a lens for the radiofrequency magnetic fields and is able to increase the penetration depth of magnetic resonance phased arrays of surface coils.

**Implications for patient care:** The device shown in the present work can increase the signal-to-noise ratio in magnetic resonance images, so that the acquisition time of the images can be reduced and thus the examination time.

**Abstract**

**Purpose:** To test the ability of a new class of passive electromagnetic device to increase the penetration depth of phased arrays of surface coils for magnetic resonance (MR) imaging systems. This new device is based on the emerging technology of metamaterials and behaves like a lens for the radiofrequency magnetic fields.

**Materials and methods:** The presented device was tested in several 1.5-T MR systems from different companies in combination with different phased arrays. One of the authors was enrolled as volunteer for the experiments. In these experiments his knees were imaged by using a dual phased array. The device was placed between the knees to check that the penetration depth of the coils was improved by this passive device.

**Results:** In all the experiments the presented device was successfully tested and it was checked that the knees of the volunteer can be imaged at deeper distances and that the signal-to-noise-ratio (SNR) in the obtained MR images was improved by the presence of the lens.

**Conclusion:** The presented device has proven to increase the penetration depth of MR phased arrays of surface coils. The lens was tested by means of the MR imaging of the knees but it can be used to image any pair of joints simultaneously by placing it between the joints. The positive results suggest the possibility of using the lens to image the female breast. This would make it possible to increase the SNR without higher fields, thus fulfilling the safety regulations governing the standard absorption rate (SAR).

# Introduction

Metamaterials are composite materials with effective permeability and permittivity determined by their structure rather than by the intrinsic properties of the material components (which are conventional conductors and dielectrics) and exhibit electromagnetic properties that are unusual in natural materials, as for example the possibility of showing both negative permeability and permittivity simultaneously (1). One of the most interesting applications of metamaterials is the fabrication of lenses that go beyond the diffraction limit of classic optics. As suggested theoretically by Pendry in his famous paper (2), a metamaterial slab with negative effective permeability and/or permittivity can function as a lens which would be able to image an electromagnetic source with a resolution smaller than the wavelength of the electromagnetic fields. Since metamaterials are periodic structures fabricated by means of the repetition of a resonant element, the interesting properties occur in a very narrow band of frequencies due to the resonant nature of its constituent elements. This narrow bandwidth is the main drawback for some of the applications of metamaterials, for example in the design of an invisibility shell at optical frequencies (3), which is an application of metamaterials that has received great attention in news media. However, the narrow bandwidth is not a problem for an application such as magnetic resonance (MR) imaging since images are acquired by measuring radiofrequency (RF) signals inside a relatively narrow bandwidth of a few tens of kilohertz. In addition, since the wavelength associated with MR fields is of the order of the meter, it is possible to use conventional fabrication circuit techniques to make composite materials where the size of constituent elements of the periodic structure is much more smaller than the wavelength. For the abovementioned reasons, new metamaterial technology could be useful in MR imaging.

The application of metamaterials in MR imaging has been explored in a few works (4-8). In all these studies, the chosen metamaterial components are non-ferromagnetic so they do not interact with the static magnetic field. Nevertheless, due to the internal structure of the metamaterials they can interact with the RF magnetic field. A type of resonant element called "Swiss roll" was used for the metamaterials analyzed in (4-7). The "Swiss roll" consists of a conductive layer which is wound on a spiral path around a cylinder with an insulator separating consecutive turns. In (8), capacitively-loaded copper loops were used as constituent elements. In (4) and (5) the reported experiments prove that "Swiss-roll" metamaterials can guide the RF flux from a sample to a remote coil when the metamaterial is placed between the sample and the coil. The guiding behaviour is due to the high effective permeability of the metamaterial. The reported experiments also prove that these metamaterial guides can be employed in imaging (4) and spectroscopic (5) experiments for excitation as well as reception. However, the experiment in (5) shows that the signal-to-noise ratio (SNR) measured with the coil placed directly on the sample is always higher than the SNR measured with the flux guide being placed between the sample and the coil. This is due to the inherent losses in the "Swiss rolls" and it prevents the use of such guides to improve the SNR of a conventional coil. The authors of the present paper showed a metamaterial device in (8) that did not work as a guide with a high effective permeability but behaved like a lens that collimated the radiofrequency magnetic field due to an effective permeability equal to -1, that is, the permeability of vacuum with opposite sign. In the experiment reported in (8), our lens was employed to increase the penetration depth of a standard surface coil due to the ability of the lens to collimate the RF flux coming from deep tissues. In this experiment, a three-inch surface coil was first used to image one knee of one of the authors. Then, the lens was placed between both knees and thus the surface coil was able to image both knees simultaneously. Since one of the knees was between the surface coil and the lens, the main source of noise for the coil was this knee, so the inherent losses in the lens did not degrade the SNR. In general, the lens could be used to image pairs of joints simultaneously by using only one coil. A more promising application is for imaging of the female breast. Increasing the SNR by using higher field scanners is limited by the standard absorption rate (SAR), the rate at which RF energy is absorbed by the body. Metamaterial lenses placed in between the breasts and combined with a breast-dedicated phased array coil would make it possible to increase the SNR without higher fields. In our research toward this goal, in this work we have experimentally checked how the lens can increase the SNR of the images obtained with dual phased arrays.

**Materials and methods**

The lens reported in (8) consisted of a three-dimensional array of capacitively-loaded copper rings with 18x18x2 unit cells, each cell being 1.5 cm, and the with a total device length (XY cm) and width of (XY cm) 27 cm and a thickness of 3 cm. In the present work, a similar lens was used but with a three unit cell depth and a thickness of 4.5 cm. The ability of the lens to focalize the RF fields grows with the thickness but also ohmic losses in the lens grow with the thickness since there are more rings. In our lab we performed numerical simulations (9) and measurements and predicted that the chosen thickness of 4.5 cm was the optimum value. The lens was designed to work at the Larmor frequency corresponding to a field of 1.5 T and it was successfully tested in several 1.5T scanners from SIEMENS, GENERAL ELECTRIC and PHILIPS, in similar experiments. The details of the test carried out in the Siemens scanner are described below. A bilateral four channel phased array coil designed for bilateral proton imaging of the carotids bifurcation (Machnet BV; The Netherlands) was used in a 1.5-T system (Avanto; Siemens Medical Solutions, Erlangen, Germany) in the University Hospital Charleroi, Belgium. Axial T1-weighted images were obtained. The dynamic series consisted of a T1-weighted spin-echo sequence with repetition time 450 ms, echo time 12 ms and slices of 6 mm thickness, a field of view of 189x319 mm2 and a 228x384 matrix.

**Results**

Figure 1 is a photograph of the experimental setup with the lens being between the knees of one of the authors and with the coils placed on the knees. Figure 2 shows a sketch of the experimental setup when the lens is present and when the lens is absent and the T1-weighted images corresponding to both cases. When the lens is present both knees are much more visible and a quantitative increase in the SNR of 40% was measured.

**Discussion**

A metamaterial lens for radiofrequency magnetic fields was successfully tested in several 1.5-T MRI scanners from different companies in combination with different dual phased arrays. The arrays were used to image pairs of joints simultaneously. When the lens is placed between the joints, the penetration depth of the elements of the array is increased. This lens is a completely passive device that can be designed to operate for any strength of the magnetic field. These results encourage us to investigate the possibility of using the lens to image the female breast. The lens would be placed between the breasts in a similar way as it has been done with the joints and would work in combination with dedicated-breast array coils. The lens would allow an increase in the SNR (and therefore the spatial resolution) in breast imaging. Until now, people have pursued higher field strengths to improve breast diagnostic capabilities. However, due to the risk of generating hotspots during MR imaging with higher fields, SAR limits are being enforced by the U.S. Food and Drug Administration and other healthcare safety groups and only the development of more advanced coil concepts is permitted to address the issue. Thus, the lens could provide the way to increase the SNR and the spatial resolution by fulfilling the safety regulations governing SAR. **The authors feel that, in general, the emerging technology of metamaterials could help to improve several aspects of MR imaging by providing new concepts for the advancement of the MR technology.**

## Acknowledgements

This work has been supported by the Spanish Ministerio de Educación y Ciencia under Project No. TEC2007-68013-C02-01/TCM and by the Spanish Junta de Andalucía under Project No. P06-TIC-01368. We want to thank Dr. Bavo van Riet, MR Regional Business Manager for South-West Europe from SIEMENS Medical Solutions and his partners Roger Demeure and Pierre Foucart, for providing the MRI facilities used in this work.

# Captions to Figures

Figure 1: Photograph showing the lens employed in the experiments. The lens is placed between the knees of one of the authors and a three-inch surface coil belonging to a dual phased array is visible. The photograph corresponds to the test carried out in Centro PET Cartuja, Seville, Spain, with a 1.5-T system (Signa; General Electric Medical Systems, Milwaukee, Wis).

Figure 2: Sketch of the experiment carried out in different 1.5-T systems using different phased arrays, and MR images (axial T1-weighted spin-echo sequences with repetition time 450 ms, echo time 12 ms) obtained in a system from SIEMENS (Avanto; Siemens Medical Solutions, Erlangen, Germany) in the University Hospital Charleroi, Belgium, bilateral four channel phased array coil (Machnet BV; The Netherlands). The knees are placed between the elements of the array and the distance between the knees is equal to the thickness of the lens (4.5 cm). (a) As expected the signal is stronger near the surface coil and more superficial structures are better represented than deeper ones due to the longer distance to the coil. (b) The lens is placed between the knees and then medial structures such as the vastus medialis and sartorius muscles in both legs are more clearly demonstrated compared to current methods.

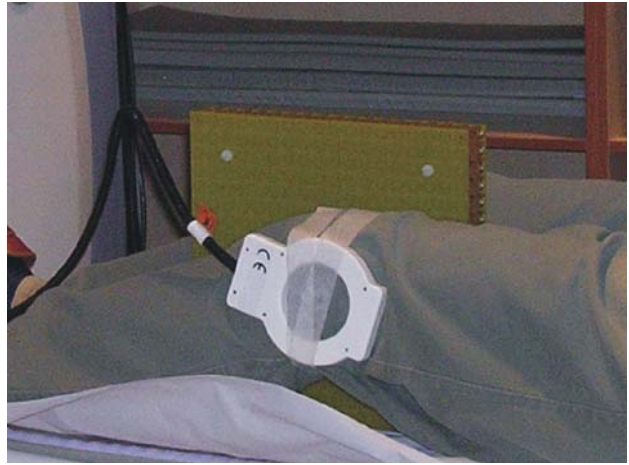

Figure 1

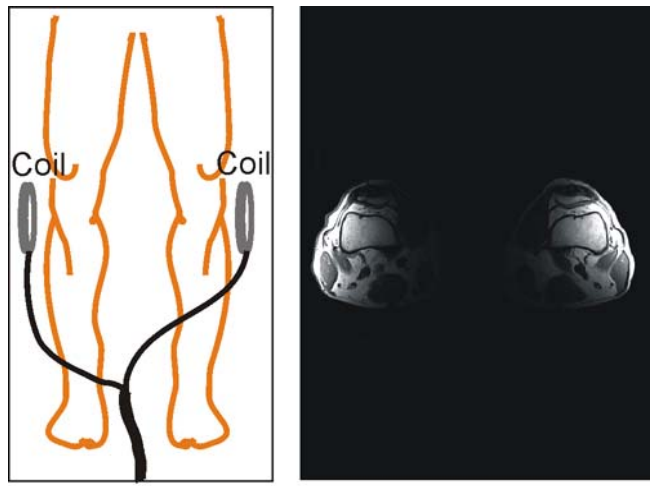

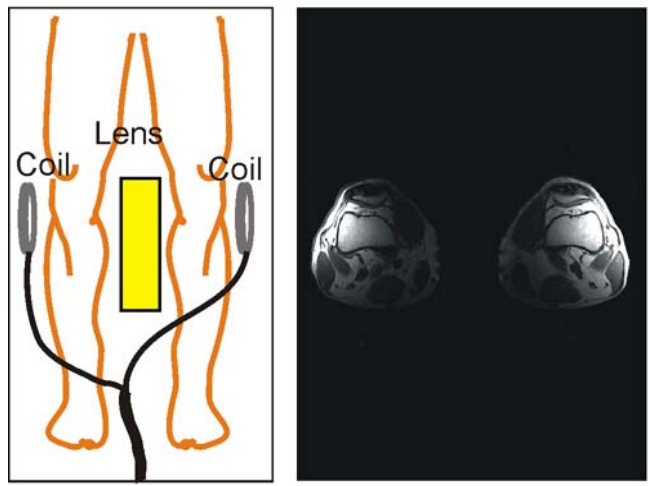

Figure 2